\documentclass[a4paper]{article}
\usepackage{geometry}
\usepackage[utf8]{inputenc}
\usepackage[T1]{fontenc}

\usepackage{url}
\usepackage{cite}
\usepackage{amsmath,amssymb,amsfonts}
\usepackage{algorithmic}
\usepackage{graphicx}
\usepackage{textcomp}
\usepackage{xcolor}

\usepackage{comment}

\newcommand{\rtt}{\mathrm{RTT}}
\newcommand{\delay}{\mathrm{delay}}
\newcommand{\find}{\mathrm{find}}

\begin{document}

\title{Evaluation and Ranking of Replica Deployments in Geographic State Machine Replication\thanks{This work was supported by JSPS KAKENHI Grant Numbers JP16K16035 and JP18K18029.\\ \indent \textcopyright\ 2021 IEEE. Personal use of this material is permitted. Permission from IEEE must be obtained for all other uses, in any current or future media, including reprinting/republishing this material for advertising or promotional purposes, creating new collective works, for resale or redistribution to servers or lists, or reuse of any copyrighted component of this work in other works.\\ \indent $^\dagger$Corresponding author: junya[at]imc.tut.ac.jp}}

\author{Shota Numakura, Junya Nakamura$^\dagger$, and Ren Ohmura}

\date{Toyohashi University of Technology, Japan}

\maketitle

\begin{abstract}
Geographic state machine replication (SMR) is a replication method in which replicas of a service are located on multiple continents to improve the fault tolerance of a general service.
Nowadays, geographic SMR is easily realized using public cloud services; SMR provides extraordinary resilience against catastrophic disasters.
Previous studies have revealed that the geographic distribution of the replicas has a significant influence on the performance of the geographic SMR; however, the optimal way for a system integrator to deploy replicas remains unknown.
In this paper, we propose a method to evaluate and rank replica deployments to assist a system integrator in deciding a final replica deployment.
In the method, we also propose a novel evaluation function that estimates a latency of SMR protocols with round-trip time (RTT).
To demonstrate the effectiveness of the proposed method, we build thousands of geographic SMRs on Amazon Web Services and present experimental results.
The results show that the proposed method that estimates a latency based on RTTs can generate consistent rankings with reasonable calculation time.
\end{abstract}

\section{Introduction}
Services running on the client-server model may crash or behave unintentionally from time to time due to software bugs or attacks by malicious users.
To prevent such problems and continuously provide services to clients, fault tolerance is an important consideration.
\emph{State machine replication} (SMR) \cite{Schneider1990} is commonly used to improve fault tolerance by replicating a service over multiple replicas.
In SMR, the replicated service is called replicas, and the state of all replicas are kept consistent by executing a replication protocol.
Hence, using this method, an active operation can be continued as a whole even if a failure occurs in a part of the replicas.
Several SMR protocols have been proposed in previous studies \cite{Moniz2011,Cachin2001,Castro2002,Nakamura2014a,Kotla2007,Sousa2012,Bessani2014}.

An SMR that deploys replicas on a continental scale is called \emph{geographic SMR} \cite{Sousa2015,Liu2017,Eischer2018,Mao2008,Veronese2010,Coelho2018}.
Replicas in geographic SMR are separated by a large distance to withstand a catastrophic disaster, such as an earthquake.
If some of the replicas fail, the service can be continued by the replicas in other \emph{sites} (regions).
With the development of public cloud services after the 2000s, geographic SMR can be easily realized.

Although geographic SMR could have been easily implemented, ways of obtaining the best performance using the optimal replica deployment remain unclear.
Performance of a replica deployment depends on several factors, including the location of the leader replica, distances between replicas, and distances between clients and replicas.
For example, if replicas are deployed in nearby regions, the time taken for request processing can be shortened, but the fault tolerance will be reduced.
In contrast, if the replicas are distributed farther apart from one another, the fault tolerance will increase, but the processing time for a normal request will be slower.

In this paper, we propose a performance-estimation method to determine the optimal replica deployment for building a service using geographic SMR.
First, we define the task to find the optimal replica deployment among all possible candidates as \emph{replica deployment decision problem}, which requires to output a ranking of all possible replica deployments sorted by their latencies.
The proposed method solves this problem by using an evaluation function that estimates a latency of each replica deployment based on the \emph{round-trip time} (RTT), which is generally regarded as an important parameter in geographic SMR.
Although it is unrealistic to actually build all possible replica deployments and measure their latencies, RTTs can be measured relatively easily.
Therefore, this evaluation function is practical and can be used to select the optimal deployment for actual service construction.

Finally, we conduct an experimental evaluation using Amazon Web Services with 15 regions to demonstrate the effectiveness and practicality of the proposed method.
In the experiment, we actually build thousands of geographic replications and measure their latencies; then we create the measured latency ranking and compare it against the rankings generated by the proposed method.
The results exhibit that the proposed method with the RTT-based evaluation function can generate a consistent ranking with reasonable calculation time.

In particular, this paper makes the following contributions:
\begin{enumerate}
\item It presents a new method that generate a ranking to assist deciding a replica deployment for geographic SMR.
\item It also presents a evaluation function that consistently calculates latency of a replica deployment by using round-trip time between sites, which can be easily measured compared with the actual latency of the deployment.
\item It conducts exhaustive experiments with thousands of replications built on Amazon Web Services, and evaluates the proposed method and the evaluation function.
\end{enumerate}

\section{Background}
\label{sec:backgrund}

\subsection{State Machine Replication}
\label{sec:smr}

\emph{State machine replication} (SMR) \cite{Schneider1990} is a replication method for the client-server model.
In SMR, the server is modeled by a state machine; thus, on receipt of a message, the server changes its state and sends messages to other processes if necessary.
The server's role is replicated over $n$ replicas that independently operate the functions on distinct hosts and interact with clients via request and response messages.

Client requests to be executed are submitted to all replicas, and the order in which different replicas receive these requests may differ due to variations in the communication delays.
Therefore, the replicas execute a replication protocol to guarantee that they process requests in the same order to maintain consistency.
After a replica processes a request, it replies to the client with the execution result.

There are two variations of SMR;
SMR that can withstand crash failures (resp. Byzantine failures) is called CFT SMR (resp. BFT SMR).
The number of faulty replicas that a replication can tolerate $f$ is related to $n$ as follows \cite{Lamport2002}:
$n \geq 2f +1$ for CFT SMR and $n \geq 3f +1$ for BFT SMR.
Hereafter, we assume BFT SMR and $n = 4$ (i.e., $f=1$); however, the proposed method is applicable for any $n$ and $f$ of BFT SMR and CFT SMR.

\subsection{Related Work}
\label{sec:relatedwork}

The problem of determining the optimal replica deployment has been extensively studied in the field of data replication.
Cook et al. formulated the time required to read and write data as a cost in a simple read-write policy (when reading a data object, refer to one replica. When writing data, a client transfer the data to all servers that have its replica) and proved that this problem is NP-complete \cite{Cook2002}.
They also proposed an approximation algorithm for the problem.
Although the target replication problem is different, their formulation is very similar to the evaluation function proposed in this paper.
The survey by Sen et al. \cite{Sen2015} provides a comprehensive overview of the previous studies on the data location optimization problem using mathematical models.

In the field of geographic SMR, there are a few methods that optimize a replica deployment \cite{Liu2017,Eischer2018}.
In \cite{Liu2017}, Liu and Vukoli\'{c} proposed two methods for geographic SMR: Droppy that dynamically relocates a set of replication leaders according to given replication settings and workload situations, and Dripple that divides the replicated system state into multiple partitions so that Droppy can efficiently relocate the leaders.
Eischer and Distler proposed Archer \cite{Eischer2018} that relocates leaders based on their response times as measured by clients.
A Hash-chain-based technique was employed in the protocol to allow clients to detect illegal phases caused by Byzantine replicas to prevent such replicas from being wrongly assigned as leaders.

In this paper, we propose a method that can help identify the best replica deployment when building an geographic SMR.
The proposed method differs from these prior studies in several ways.
First, the proposed method can be used with any replication protocol by defining an evaluation function to calculate the estimated latency of different replica deployments.
In contrast, although Droppy and Archer can dynamically relocates the leader replica locations, they only support leader-based replication protocols.
Second, the proposed method can also identify the best replica deployment from all possible replica deployments; this complements these existing methods, which are limited to determining an assignment of replication roles to the replicas in a replication.

\section{Replica Deployment Decision Problem}
\label{sec:problem-definition}

We formally define the problem addressed herein as a \emph{replica deployment decision problem}.
In the definition, we call a location wherein a replica (or a client) can be deployed to as a \emph{site}\footnote{For example, if the SMR is built on a public cloud service, each region is a site; if it is built in facilities on premises, each data center is a site.}.
In the problem, the following inputs are provided by a user.
\begin{itemize}
    \item $n$: the number of replicas that the user wants to deploy
    \item $SC$: a set of candidate sites wherein replicas can be deployed
    \item $C$: a set of client locations
\end{itemize}

The goal of this problem is to output a ranking\footnote{
The proposed method outputs not only the best replica deployment, but also the whole ranking of all possible deployments, because the best deployment may not be acceptable for some reason other than latency.
}
of replica deployments sorted by latency (of course, a replica deployment with smaller latency is ranked higher).
The user will then choose the final replica deployment for the SMR from this ranking.
Here, latency is defined as the time taken by a client from sending a request to the replicas until receiving its response.

\section{Proposed Method}
\label{sec:proposed-method}

In this section, we propose a method to solve the replica deployment decision problem and to determine the optimal replica deployment from all the possible deployments for geographic SMR.
Using the proposed method, any replication configuration can be evaluated without actually building it.

\subsection{Overview}
\label{sec:proposed-method-overview}

Figure \ref{fig:approach} illustrates the overview of the proposed method and the method consists of the following steps:
\begin{figure}[tp]
    \centering
    \includegraphics[width=65mm]{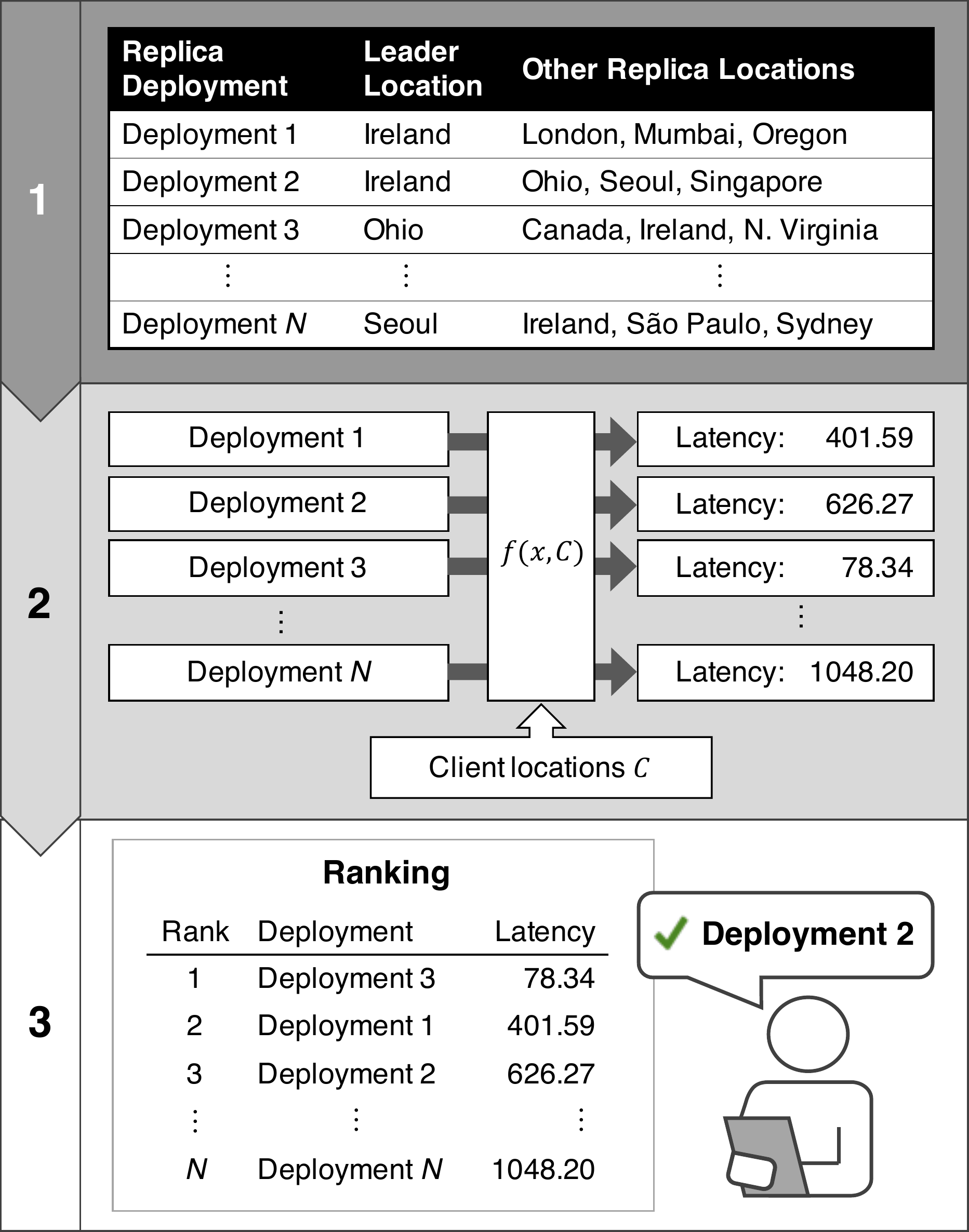}
    \caption{Overview of the proposed method}
    \label{fig:approach}
\end{figure}
\begin{enumerate}
\item First, a set, $DC$, of all possible replica deployments is created based on $SC$ and $n$.
Each replica deployment is expressed as a pair of locations for the leader and the other replicas\footnote{Here, we assume rotating coordinator-based SMR protocols similar to those \cite{Lamport1998,Castro2002,Sousa2012,Kotla2007}.
If the proposed method is applied to leader-less SMR protocols similar to those \cite{Moniz2011,Cachin2001,Nakamura2014a}, then each replica deployment is simply expressed as a set of replica locations of size $n$.
}.
\item Next, for each replica deployment $x \in DC$, its latency is estimated using the evaluation function $f(x, C)$ based on the measured RTTs.
This function is further described in Section \ref{sec:evaluation-function}.
\item The elements in $DC$ are sorted based on their calculated latency; the sorted result is outputted as the ranking for the inputs.
\end{enumerate}
Thus, the replica deployment with the shortest latency is ranked as the best replica deployment.

\subsection{Evaluation Function $f(x, C)$} 
\label{sec:evaluation-function}

The evaluation function $f(x, C)$ outputs an estimated latency based on replica deployment, $x$, and the client locations, $C$ by tracing message transmissions specific to a replication protocol being used.
The function plays an important role in the proposed method.

\subsubsection{Approach}
\label{subsubsec:evaluation-function-approach}

If site candidates $SC$ is large, it is impractical to actually build SMRs with all possible replica deployments to evaluate their latencies.
Therefore, the evaluation function estimates them based on round-trip time (RTT) between sites, which can be measured more easily, and outputs as an latency for that deployment.
In other words, before using the proposed method, a user must measure RTTs between candidate sites in advance.
Here, the time required for message processing in a replica is disregarded because the communication delay between replicas is relatively large compared with the processing time in a geographic SMR.

Assuming that a latency can only be estimated from the communication time, two factors must be considered: the types of message communications (i.e., \emph{message transmission patterns}) that constitute latency and the communication time between sites.
The message transmission pattern can be found by referring to an SMR protocol used in a replication.
Then, for a given set $C$ of clients and replica locations $x$, the function simulate the transmission and receipt of messages based on the message transmission pattern of the replication protocol and the measured RTTs.

\begin{figure}[tp]
    \centering
    \includegraphics[width=65mm]{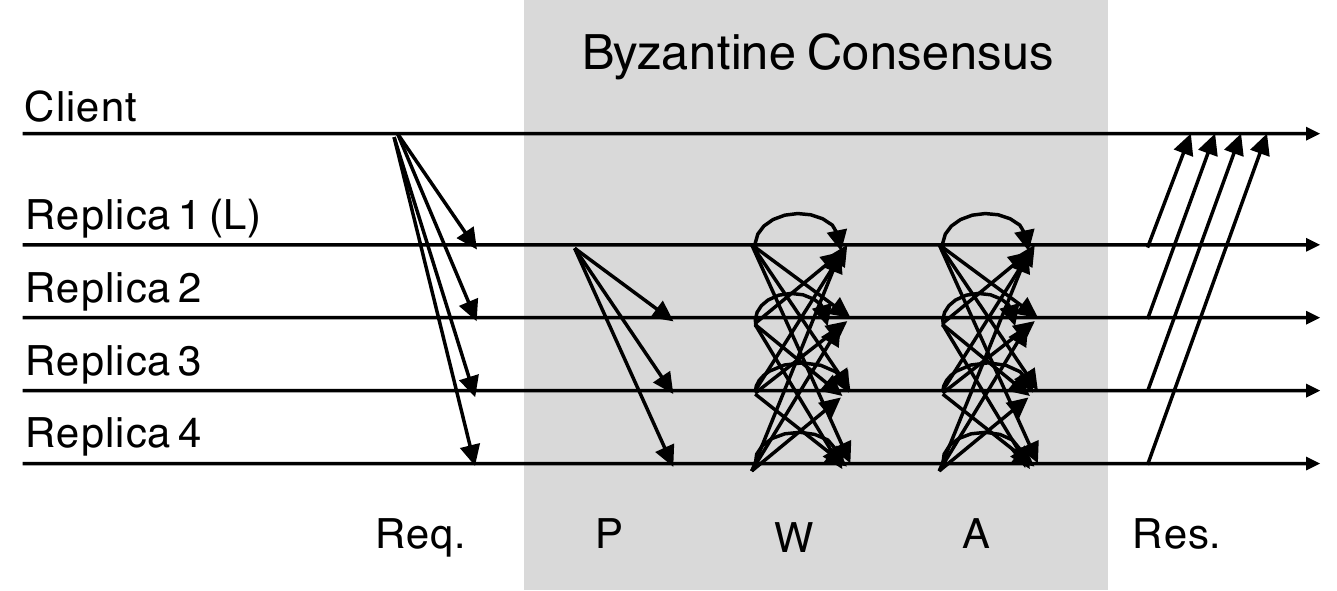}
    \caption{The message transmission pattern for Mod-SMaRt protocol \cite{Sousa2012} in BFT-SMaRt \cite{Bessani2014}.
        Replica 1 is the leader replica and Req., P, W, A, and Res., indicate Request, Propose, Write, Accept, and Response messages, respectively.}
    \label{fig:bft-smart-message-flow}
\end{figure}

Here, we model the message transmission pattern of Mod-SMaRt \cite{Sousa2012} of BFT-SMaRt \cite{Bessani2014} as an example; however, we believe the same approach can be applied to other SMR protocols.
In Mod-SMaRt, a special replica (called a \emph{leader} replica) determines the order in which requests are executed and communicates this order to the other replicas.
The message transmission pattern involves five types of messages that are exchanged among the client and replicas to process the request, as shown in Fig. \ref{fig:bft-smart-message-flow}.
First, the client sends a request to each replica (Request).
When the leader replica receives the request, it sends Propose messages to each replica to propose a candidate value for agreement (Propose).
Then, Write and Accept messages are exchanged between all replicas to confirm the validity of the candidate values to determine the final agreed value (Write and Accept).
Finally, the replicas execute the ordered request and return the result to the client (Response).
Hereafter, an RTT and a message transmission delay between sites $a$ and $b$ is denoted as $\rtt(a, b)$ and $\delay(a, b) = \rtt(a, b)/2$, respectively.

\subsubsection{Latency Formulation}
\label{subsubsec:latency-calculating}

The evaluation function $f$ estimates a latency of each client location $c \in C$, and outputs the average of these latencies as follows:
\begin{equation}
    f(x, C) = \sum_{c \in C} f_c(x, c) / |C|,
\end{equation}
where $f_c$ is a evaluation function for a single client.
Hereafter, we explain how $f_c(x, c)$ calculates a latency on a replica deployment.
The message pattern of Mod-SMaRt comprises five parts as depicted in Fig.~\ref{fig:bft-smart-message-flow}, and we denote the timings of these parts by $S_{req}$, $S_{pro}$, $S_{wrt}$, $S_{acc}$, and $S_{res}$, respectively. 
If necessary, we denote the timing for a specific replica $r_i$ by adding a superscript such as $S_{pro}^i$.

First, we calculate the timing $S_{req}$ at which the leader receives a request.
In the replication protocol, a request message is sent from a client to each replica although only the leader replica processes the request in the fault-free case;
thus, $S_{req}$ can be expressed as the average of the RTTs from each client $c$ to the leader replica $l$:
\begin{equation}
    S_{req} = \sum_{c \in C} \delay(c, l) / |C|.
\end{equation}

Then, the leader sends the request to each replica as Propose messages; the timing $S_{pro}^i$ at which the replica $r_i$ receives the Propose message is expressed as follows:
\begin{equation}
    S_{pro}^i = S_{req} + \delay(l, r_i).
\end{equation}

When a replica receives the Propose message, it broadcasts a Write message to all replicas.
Each replica accepts the Write message when it receives the same Write messages from a majority $\lceil (n+1)/2 \rceil$ of the replicas.
The timing $S_{wrt}^i$ at which replica $r_i$ accepts the Write messages can be calculated based on the timing at which the replica $r_{i}$ receives the Write message sent from replica $r_j$:
\begin{equation}
    S_{wrt}^i = \find(T^i_{wrt}, \lceil (n+1)/2 \rceil),
\end{equation}
where $t_{wrt}(r_i, r_j) = S_{pro}^j + \delay(r_j, r_i)$, $T_{wrt}^i = \{ t \mid t_{wrt}(r_i, r_j), 0 \leq j < n \}$, and $\find (S, k)$ is a function that returns the $k$-th smallest element of set $S$.

An Accept message is sent in the same way as Write messages.
Therefore, if we define $t_{acc}(r_i, r_j) = S_{wrt}^i + \delay(r_j, r_i)$, $S_{acc}^i$ is
\begin{equation}
    S_{acc}^i = \find(T^i_{acc}, \lceil (n+1)/2 \rceil),
\end{equation}
where $T_{acc}^i = \{ t \mid t_{acc}(r_i, r_j), 0 \leq j < n \}$.

Finally, when a replica receives a majority of Accept messages, it executes the request and sends the execution result to the client as a Response message.
When a client receives the same response message from $f + 1$ distinct replicas, it accepts the result.
Therefore,
\begin{equation}
    f_c(x, c) = S_{res} = \find(T_{res}, f+1),
\end{equation}
where $T_{res} = \{ t \mid S_{acc}^i + \delay(r_i, c), 0 \leq i < n \}$.

\section{Evaluation}
\label{sec:evaluation}

In this section, we examine the effectiveness of the proposed method described in Section \ref{sec:proposed-method}.
First, the evaluation of replica deployments in terms of the RTT is verified in Section \ref{sec:rtt-experiment}.
Next, the latencies of thousands of replica deployments on a public cloud service are measured to evaluate the accuracy of the ranking generated by the proposed method in Section \ref{sec:latency-experiment}.
Finally, Section \ref{sec:ranking-experiment} characterizes the time that it takes to generate a ranking.

All experiments are conducted using Amazon Web Services EC2, a representative public cloud service.
We use 15 regions\footnote{N.~Virginia, Ohio, N.~California, Oregon, Mumbai, Seoul, Singapore, Sydney, Tokyo, Canada Central, Frankfurt, Ireland, London, Paris, S\~{a}o Paulo} of Amazon EC2 as site candidates $SC$ for replica deployments (i.e., $|SC| = 15$).
Replica and client programs are executed on Ubuntu Server 16.04 (64 bit).
For replicas and clients, we use t2.micro instances that have one vCPU, 1 GiB memory, EBS storage, and a network interface of "Low to Moderate" performance.

\subsection{Validation of the Use of RTTs}
\label{sec:rtt-experiment}
The proposed method calculates lantecies based on the RTTs between sites.
Here, we evaluate whether it is appropriate to use the RTTs for estimating lantecy and how long the generated ranking is valid.

\subsubsection{Method}
\label{subsubsec:rtt-measuring-method}

An instance is deployed in each of the regions, and the \verb,ping, command is executed against the instances in the other 14 regions every two seconds.
RTTs were measured during the following periods (all times are displayed in UTC in 24-h notation):
\begin{itemize}
\item Term A: March 7, 19:27 -- 22:13, 2018
\item Term B: January 11, 11:14 -- January 28, 3:41, 2019
\item Term C: April 15, 15:48 -- April 23, 11:15, 2019
\end{itemize}

\subsubsection{Results and Discussion}
\label{subsubsec:rtt-results}

\begin{figure}[tp]
	\centering
	\includegraphics[scale=0.80]{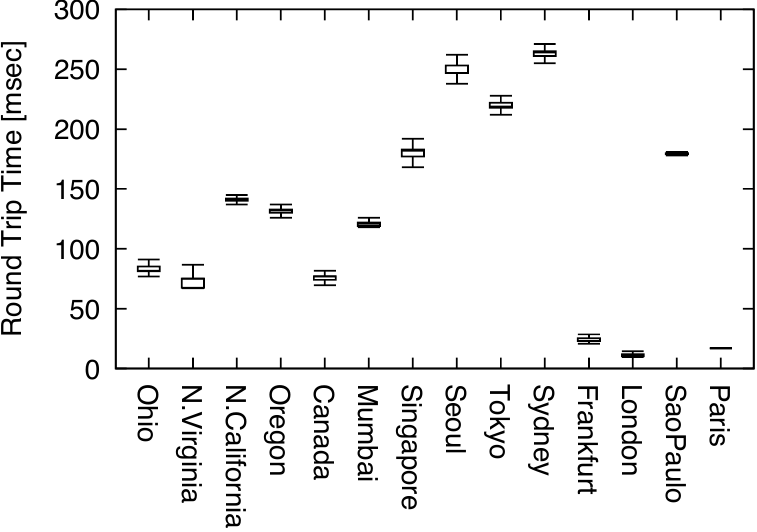}
	\caption{Distribution of RTT from Ireland to each region during term A.}
	\label{fig:rtt-ireland-all}
\end{figure}

RTTs measured during Term C are shown as a boxplot in Fig. \ref{fig:rtt-ireland-all} (only the results for the Ireland instance are shown due to space limitations).
Although RTT varied from region to region, these variations were small.
The largest variation was observed between the Ireland and Singapore regions, and its mean and standard deviation were 180.3 and 24.1 ms, respectively.

Next, we compare the RTTs from Ireland to Singapore (where the largest variations were observed) during terms A, B, and C.
The average RTTs were 175.3 ms in term A, 179.8 ms in term B, and 180.3 ms in term C.
Over the 13 months between term A and term C, RTT increased by 5 ms.
Although this may seem like a small difference, if similar changes occurred between all regions, it is likely that the ranking generated by the proposed method would change considerably.

\begin{figure}[tp]
    \centering
    \includegraphics[scale=0.31]{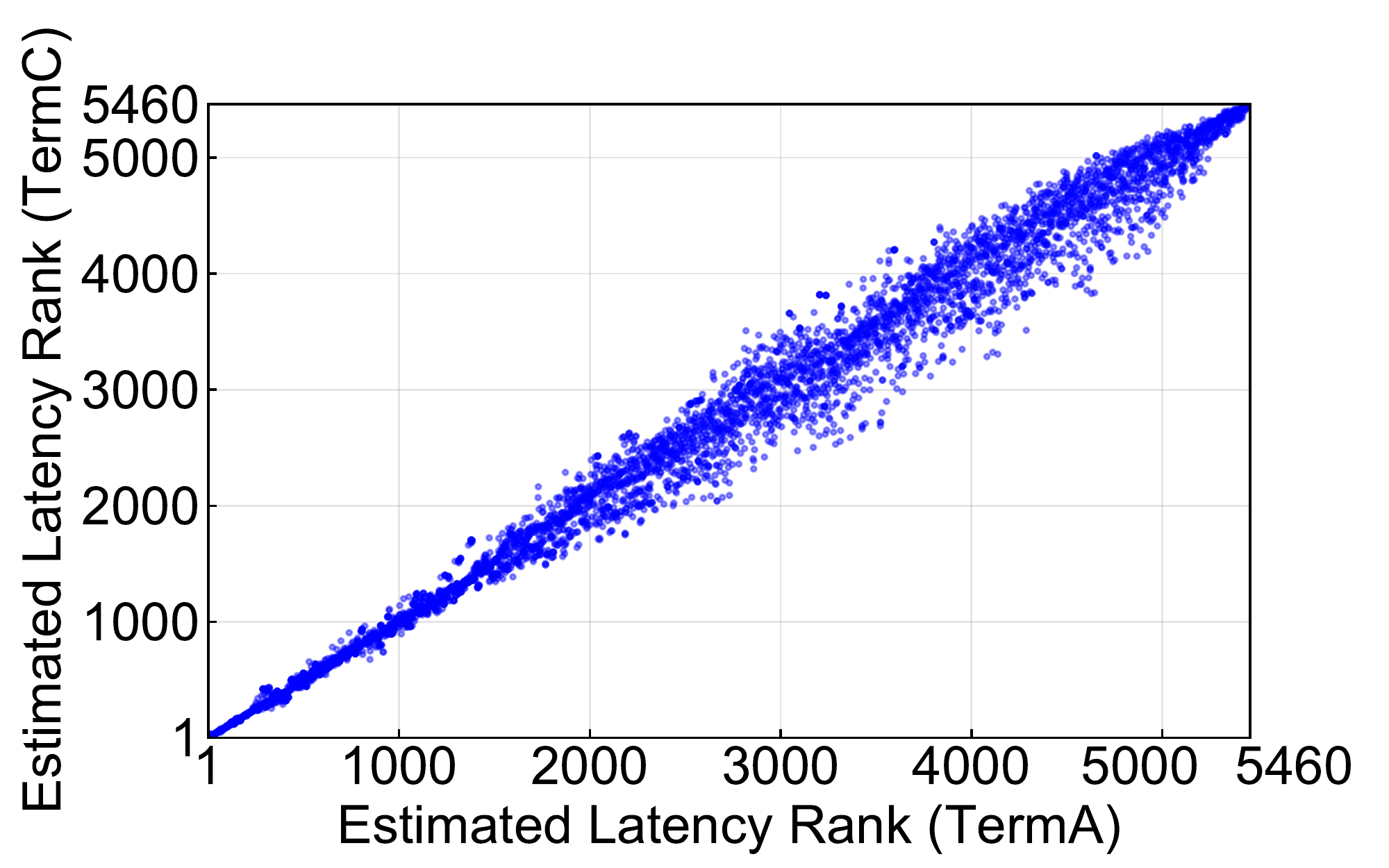}
    \caption{Difference between the rankings of Terms A and C.}
    \label{fig:rtt_termA_vs_termC}
\end{figure}

To investigate how these difference affects a replica deployment ranking, we generated two rankings from the RTTs measured during Terms A and C with the client location Multiple (see Section \ref{subsubsec:latency-measuring-method} for its definition).
Figure \ref{fig:rtt_termA_vs_termC} shows the correlation between these rankings. 
We can observe that the RTT changes affected the ranking certainly, especially for the 2000--5000 ranks.
The largest difference happened on the replica deployment of Tokyo (leader), Canada, Oregon, and Singapore.
The deployment was in 3523rd place in the term A ranking, while it was in 2688th place in the term C ranking.

The results indicate that the RTT variations in the public cloud are sufficiently small in the short term; thus, estimating a replica deployment in terms of based on the RTTs between sites is valid.
In contrast, RTTs between regions changed over long periods (on the order of one year).
Therefore, a replica deployment that is found to be optimal may no longer be optimal after a long time has passed, suggesting that replicas should be relocated periodically to maintain optimal performance.

\subsection{Ranking Accuracy}
\label{sec:latency-experiment}
Here, we discuss the accuracy of a ranking generated by the proposed method by comparing the rankings with those derived from the experimentally measured latencies of all possible replica deployments.

\subsubsection{Method}
\label{subsubsec:latency-measuring-method}

We introduce a baseline evaluation function $f_{simple}(x, C)$ to compare the accuracy of the evaluation function of the proposed method.
This function roughly estimates a latency based on a simplified message pattern for Mod-SMaRt.
First, it divides the pattern into three parts: Request, Byzantine Consensus, and Response as Fig.~\ref{fig:bft-smart-message-flow} and calculates their timings $S_{req}$, $S_{con}$, and $S_{res}$ as follows:
$S_{req}$ is the average of the half RTTs from each client to the leader replica.
$S_{con}$ is the sum of the half RTTs between all pairs of replicas.
$S_{res}$ is the average of the half RTTs from each replica to each client.
Finally, this function outputs the sum of these timings as a latency.

In this experiment, all possible replica deployments are built on AWS and the latency of each one is measured.
We do not assume that multiple replicas are deployed in the same region.
Since $|SC| = 15$ and $n = 4$, the total number of possible replica deployments $|DC| = |SC| \times {}_{|SC|-1}C_{n-1} = 5,460$.
If replicas are deployed to the same combination of regions, the location of the leader replica may differ; hence, such deployments are considered independently. 

As with replicas, it is assumed that the clients are also located in the AWS regions.
To evaluate the effects of the number and locations of clients, clients are placed in geographically distant regions, namely Ireland, Sydney, and N.~Virginia.
The case wherein multiple clients are placed in multiple regions (we call this deployment as ``Multiple'') is also evaluated: 10 clients are placed in Ireland, 3 clients are placed in Sydney, and 5 clients are placed in N.~Virginia.

SMR is built using the open-source SMR library BFT-SMaRt \cite{Bessani2014}\footnote{\url{https://github.com/bft-smart/library/releases/tag/v1.1-beta}}.
A replication is build to withstand Byzantine failures; the tolerable number of failures is $f=1$ and the number of replicas is $n=4$.
The defaults are used for all other BFT-SMaRt settings.

All latencies are measured using the sample programs LatencyClient and LatencyServer bundled in BFT-SMaRt.
LatencyClient periodically sends requests to the service and measures the latency.
LatencyServer is a dummy service that provides no functionality; it simply returns a response immediately after receiving a request from a client.
The payload sizes of the requests and responses are 1,024 bytes.
LatencyClient sends requests 50 times every 2 sec.
The top 10\% (i.e., the highest five values) and bottom 10\% (i.e., the lowest five values) of the measured values are considered as outliers and disregarded;
the average of the other values (40 values in total) is considered as the latency of the replica deployment.
The latency is estimated with the average RTTs measured during Term C in Section \ref{subsubsec:latency-measuring-method}.

\subsubsection{Results and Discussion}
\label{subsubsec:latency-results}

\begin{figure}[tp]
	\centering
	\includegraphics[scale=0.31]{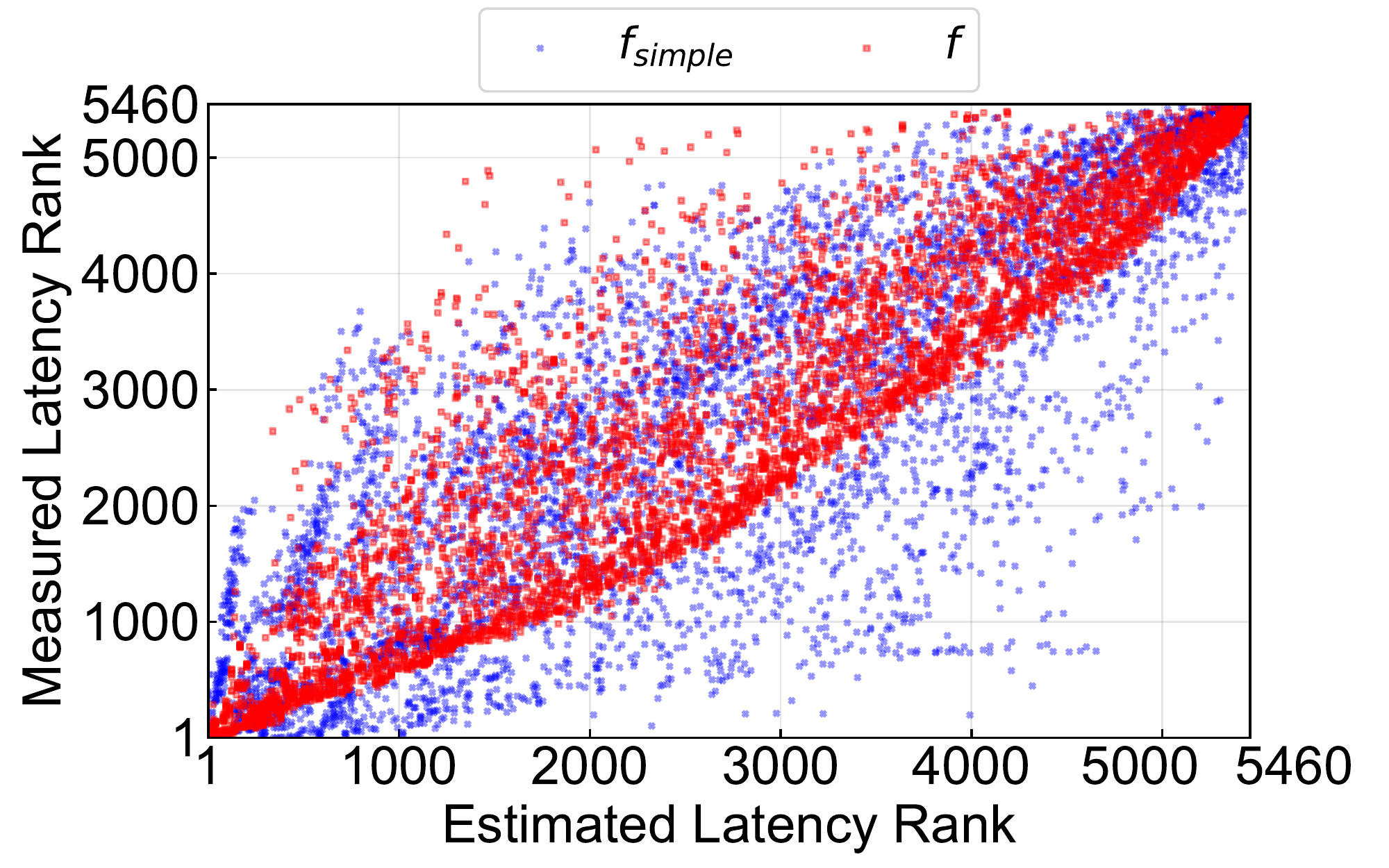}
	\caption{
    	Scatter plots of measured latency rank and estimated latency rank ($C$ = Sydney, $|DC| = 5460$).
        Each plotted point represents the latency of a replica deployment (red for $f$ and blue for $f_{simple}$).
        The horizontal axis represents the ranking derived from the latencies output by the proposed method with $f$ or $f_{simple}$, and the vertical axis represents the ranking derived from the measured latencies. 
        }
	\label{fig:multiple-5460}
\end{figure}

Figure \ref{fig:multiple-5460} shows the correlations between the rankings generated via the proposed method using the evaluation functions.
Due to space limitations, only the results for the multiple are shown.
Table \ref{tab:scatter-result-overall} also shows the root mean square error (RMSE)
calculated based on the ideal ranking (i.e., $y = x$), which perfectly matches the ranking based on the measured latencies, and the correlation coefficient (CC) for each client location.
The results indicate that the RMSE was lower and the CC was higher (exceeding 0.91 in all cases) for $f$ than for $f_{simple}$ for all client locations.
This implies that $f$ yielded more accurate rankings by tracing the communications between the replicas in detail.

\begin{table}[tp]
    \caption{RMSE and correlation coefficient (CC)}
    \begin{center}
    \begin{tabular}{|c|c|c|c|c|}
        \hline
        \textbf{} & \multicolumn{2}{|c|}{\textbf{RMSE}} &\multicolumn{2}{|c|}{\textbf{CC}} \\
        \cline{2-5} 
        \textbf{Client location} & 
        \textbf{\textit{$f_{simple}$}}& \textbf{\textit{$f$}}&
        \textbf{\textit{$f_{simple}$}}& \textbf{\textit{$f$}} \\ \hline
        Ireland & 759.411 &	620.686 & 0.884 & 0.922\\ \hline
        N. Virginia & 722.516 &	548.982 &	0.895 & 0.939\\ \hline
        Sydney & 985.598 & 638.275 & 0.804 & 0.918\\ \hline
        Multiple & 697.473 & 629.228 & 0.902 & 0.920\\ \hline
    \end{tabular}
    \label{tab:scatter-result-overall}
    \end{center}
\end{table}

These experiments confirmed that the proposed method can generate consistent rankings in various client locations.
Further, it was revealed that the rankings generated by $f$ are more accurate than those generated by $f_{simple}$ (particularly for the higher-ranked deployments).
Hence, a higher reproducibility of the replication protocol can reproduce more accurate replica deployment ranking.

\subsection{Calculation Time to Generate a Ranking}
\label{sec:ranking-experiment}

Finally, we evaluate the calculation time required to generate a ranking with the proposed method\footnote{
All the rankings were calculated by the program implemented with Python 3.6 on the following PC: Intel Core i5 7400, Windows 10 Home 64-bit.
}.
The ranking calculation times of $f_{simple}$ and $f$ were 1.88s and 10.88s, respectively for $n=4$, that is, $f_{simple}$ is about five times faster than $f$. 
This finding indicates that more time is required to calculate the estimated latency using the evaluation function and to improve the reproducibility of the communication.

Next, we investigate the influence of the size of $SC$ on the calculation time with $f$.
Table \ref{tab:ranking-each-SC} shows the resulting calculation times for different $|SC|$ values and the corresponding calculation times per replica deployment $t/|DC|$.
The result shows that as the size of $SC$ increased, the calculation time required to generate the rankings considerably increased because the total number of replica deployments $|DC|$ increases exponentially with $|SC|$.
\begin{table}[tp]
    \centering
    \caption{Calculation time by size of site candidates $SC$ ($n = 4$)}
    \label{tab:ranking-each-SC}
    \begin{tabular}{|c|c|c|c|}
        \hline
        \textbf{$|SC|$} & \textbf{Time $t$ [sec]} & \textbf{$|DC|$} & \textbf{$t/|DC|$ [msec]}\\ \hline
        15 & 12.9 & 5,460 & 2.36\\ \hline
        20 & 39.5 & 19,380 & 2.04\\ \hline
        25 & 110.5 & 50,600 & 2.18\\ \hline
        30 & 227.3 & 109,620 & 2.07\\ \hline
    \end{tabular}
\end{table}

Furthermore, the influence of the number of replicas $n$ on the calculation time with $f$.
Table \ref{tab:ranking-each-n} shows the calculation time $t$ for different $|DC|$ values and the corresponding calculation times per replica deployment, $t/|DC|$, as $n$ is varied.
The result shows that $t$ and $|DC|$ were maximized at different values of $n$ (10 and 7, respectively) because the calculation time of a replica deployment $t/|DC|$ increases as $n$ increases.

\begin{table}[tp]
    \centering
    \caption{Calculation time by the number of replicas $n$ ($|SC| = 15$)}
    \label{tab:ranking-each-n}
    \begin{tabular}{|c|c|c|c|}
        \hline
        \textbf{$n$} & \textbf{Time $t$ [sec]} & \textbf{$|DC|$} & \textbf{$t/|DC|$ [msec]}\\ \hline
        4 & 12.9 & 5,460 & 2.36\\ \hline
        7 & 429.1 & 45,045 & 9.53\\ \hline
        10 & 792.1 & 30,030 & 26.38\\ \hline
        13 & 77.7 & 1,365 & 56.90\\ \hline
    \end{tabular}
\end{table}

These measurement results reveal that the rankings for replica deployments can be calculated in several hundred seconds when the replica number and site number are relatively small.
This is considered a reasonable calculation time since a deployed SMR is typically operated for more than one year. 
In contrast, if large numbers of replicas and $SC$ are used, the calculation time becomes very high.
In such a case, some changes need to be made so that the solution is still practical, e.g., calculations latencies in parallel, discarding replica deployments that seems to be slow, and so on.

\section{Conclusion}
\label{sec:conclusion}

In this paper, we addressed on the difficulty of determining the optimal replica deployment for geographic state machine replication by proposing a novel method to generate a ranking of all possible replica deployments.
We introduced an evaluation function that estimates a latency of each replica deployment based on the RTTs between sites, which are easy to measure without actually building the deployments.
Hence, all possible replica deployments can be evaluated and ranked accordingly to determine the optimal replica deployment for geographic SMR.
We confirmed the validity of evaluating replica deployments in terms of their RTTs.
After that, we measured the latencies of thousands of replica deployments built on Amazon Web Services, and ranked the deployments accordingly.
Then, we compared this experimentally derived ranking with those rankings generated using the proposed method.
The results exhibited that the proposed method can create a ranking with sufficient accuracy in a reasonable time.

\bibliographystyle{IEEEtran}
\bibliography{collection}

\begin{thebibliography}{10}
\providecommand{\url}[1]{#1}
\csname url@samestyle\endcsname
\providecommand{\newblock}{\relax}
\providecommand{\bibinfo}[2]{#2}
\providecommand{\BIBentrySTDinterwordspacing}{\spaceskip=0pt\relax}
\providecommand{\BIBentryALTinterwordstretchfactor}{4}
\providecommand{\BIBentryALTinterwordspacing}{\spaceskip=\fontdimen2\font plus
\BIBentryALTinterwordstretchfactor\fontdimen3\font minus
  \fontdimen4\font\relax}
\providecommand{\BIBforeignlanguage}[2]{{%
\expandafter\ifx\csname l@#1\endcsname\relax
\typeout{** WARNING: IEEEtran.bst: No hyphenation pattern has been}%
\typeout{** loaded for the language `#1'. Using the pattern for}%
\typeout{** the default language instead.}%
\else
\language=\csname l@#1\endcsname
\fi
#2}}
\providecommand{\BIBdecl}{\relax}
\BIBdecl

\bibitem{Schneider1990}
F.~B. Schneider, ``Implementing fault-tolerant services using the state machine
  approach: a tutorial,'' \emph{ACM Computing Surveys}, vol.~22, no.~4, pp.
  299--319, 1990.

\bibitem{Moniz2011}
H.~Moniz, N.~F. Neves, M.~Correia, and P.~Verissimo, ``{RITAS: Services for
  Randomized Intrusion Tolerance},'' \emph{IEEE Transactions on Dependable and
  Secure Computing}, vol.~8, no.~1, pp. 122--136, 2011.

\bibitem{Cachin2001}
C.~Cachin, K.~Kursawe, F.~Petzold, and V.~Shoup, ``Secure and efficient
  asynchronous broadcast protocols,'' in \emph{CRYPTO}, 2001.

\bibitem{Castro2002}
M.~Castro and B.~Liskov, ``Practical byzantine fault tolerance and proactive
  recovery,'' \emph{ACM Transactions on Computer Systems}, vol.~20, no.~4, pp.
  398--461, 2002.

\bibitem{Nakamura2014a}
J.~Nakamura, T.~Araragi, S.~Masuyama, and T.~Masuzawa, ``{Efficient Randomized
  Byzantine Fault-Tolerant Replication based on Special Valued Coin Tossing},''
  \emph{IEICE Transactions on Information and Systems}, vol. E97-D, no.~2, pp.
  231--244, 2014.

\bibitem{Kotla2007}
R.~Kotla, L.~Alvisi, M.~Dahlin, A.~Clement, and E.~Wong, ``Zyzzyva: speculative
  byzantine fault tolerance,'' in \emph{SOSP}, 2007.

\bibitem{Sousa2012}
J.~Sousa and A.~Bessani, ``{From Byzantine Consensus to BFT State Machine
  Replication: A Latency-Optimal Transformation},'' in \emph{EDCC}, 2012.

\bibitem{Bessani2014}
A.~Bessani, J.~a. Sousa, and E.~E.~P. Alchieri, ``{State Machine Replication
  for the Masses with BFT-SMaRt},'' in \emph{DSN}, 2014.

\bibitem{Sousa2015}
J.~Sousa and A.~Bessani, ``{Separating the WHEAT from the Chaff: An Empirical
  Design for Geo-Replicated State Machines},'' in \emph{SRDS}, 2015.

\bibitem{Liu2017}
S.~Liu and M.~Vukolic, ``Leader set selection for low-latency geo-replicated
  state machine,'' \emph{{IEEE} Transactions on Parallel and Distributed
  Systems}, vol.~28, no.~7, pp. 1933--1946, 2017.

\bibitem{Eischer2018}
M.~Eischer and T.~Distler, ``{Latency-Aware Leader Selection for Geo-Replicated
  Byzantine Fault-Tolerant Systems},'' in \emph{DSN-W}, 2018.

\bibitem{Mao2008}
Y.~Mao, F.~P. Junqueira, and K.~Marzullo, ``Mencius: Building efficient
  replicated state machines for wans,'' in \emph{OSDI}, 2008.

\bibitem{Veronese2010}
G.~S. Veronese, M.~Correia, A.~N. Bessani, and L.~C. Lung, ``{EBAWA}: Efficient
  byzantine agreement for wide-area networks,'' in \emph{HASE}, 2010.

\bibitem{Coelho2018}
P.~Coelho and F.~Pedone, ``Geographic state machine replication,'' in
  \emph{SRDS}, 2018.

\bibitem{Lamport2002}
L.~Lamport, ``Lower bounds for asynchronous consensus,''
  \url{https://lamport.azurewebsites.net/pubs/bertinoro.pdf}, 2002, [Online;
  accessed July 31, 2019].

\bibitem{Cook2002}
S.~A. Cook, J.~Pachl, and I.~S. Pressman, ``The optimal location of replicas in
  a network using a {READ}-{ONE}-{WRITE}-{ALL} policy,'' \emph{Distributed
  Computing}, vol.~15, no.~1, pp. 57--66, 2002.

\bibitem{Sen2015}
G.~Sen, M.~Krishnamoorthy, N.~Rangaraj, and V.~Narayanan, ``Facility location
  models to locate data in information networks: a literature review,''
  \emph{Annals of Operations Research}, vol. 246, no. 1-2, pp. 313--348, 2015.

\bibitem{Lamport1998}
L.~Lamport, ``The part-time parliament,'' \emph{{ACM} Transactions on Computer
  Systems}, vol.~16, no.~2, pp. 133--169, 1998.

\end{thebibliography}

\end{document}